\documentclass{ptapap}

\author{Mónica Taormina}[CAMK]
\author{Bogumił Pilecki}[CAMK]
\author{Radosław Smolec}[CAMK]
\affil[CAMK]{Nicolaus Copernicus Astronomical Center\\
  Bartycka 18, 00--716 Warszawa, Poland}

\title{Pulsation Theory Models for Cepheids in Eclipsing Binary Systems}

\begin{document}

\maketitle

\begin{abstract}
Physical parameters were recently measured for several Cepheids in eclipsing binary systems in the LMC. It is a good opportunity to compare these results with models obtained from pulsation theory. Having well determined physical parameters of the stars, we can calculate the corresponding pulsation periods for a given mode and check for instability. For stars that lack data, unknown or poorly defined parameters can be further constrained. We present the comparison for six Cepheids, showing that, in general, the results are in very good agreement.

\end{abstract}

Because of their famous period-luminosity relation, Cepheids play a very important role in distance determination. They are also key objects for testing the predictions of stellar evolution and stellar pulsation theory. 
In the last decade many Cepheids in eclipsing binary systems were found, giving us the possibility to determine their physical parameters (like mass, radius) with a very good precision, when the light and radial velocity curves are analyzed together. However in some cases, when only one eclipse per cycle is visible \citep[e.g.][]{pilecki:2015}, or a Cepheid is a member of a single-lined system \citep{pilecki:2017}, we need additional information to solve the system. 
\cite{pilecki:2017} showed that in this case, we can use stellar pulsation theory to obtain necessary constraints for the solution. Here we apply this method to all the analyzed systems with classical Cepheids and try to improve the solutions for which fundamental parameters were not precisely determined.

\begin{figure}
  \centering
  \begin{minipage}{0.48\textwidth}
    \includegraphics[width=\textwidth]{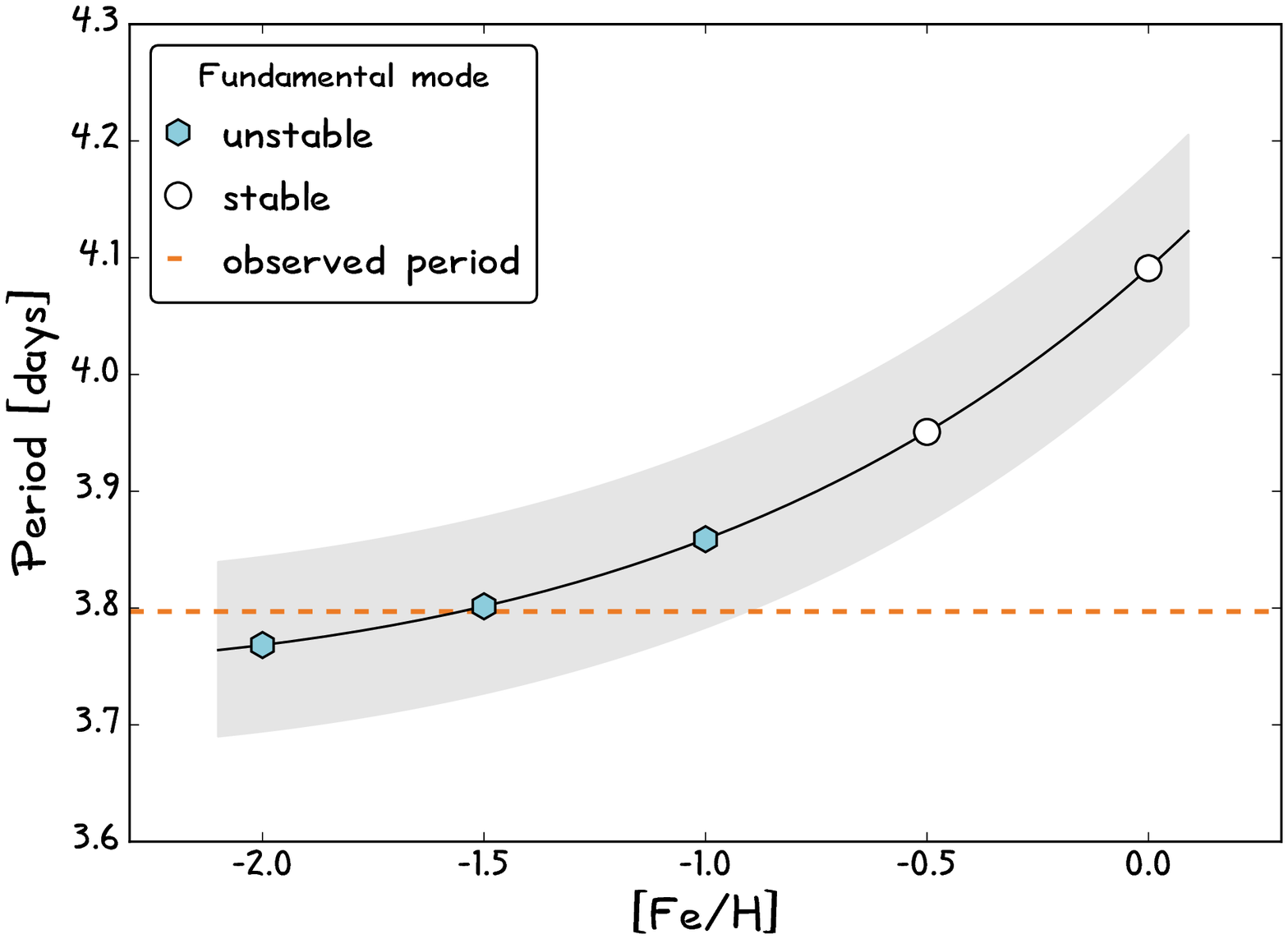}
    \caption{Periods calculated for each metallicity for LMC-CEP-0227. The uncertainty in the modelling of the pulsation period is represented with gray band.
    }
    \label{fig:fig1}
  \end{minipage}
  \quad
  \begin{minipage}{0.48\textwidth}
    \includegraphics[width=\textwidth]{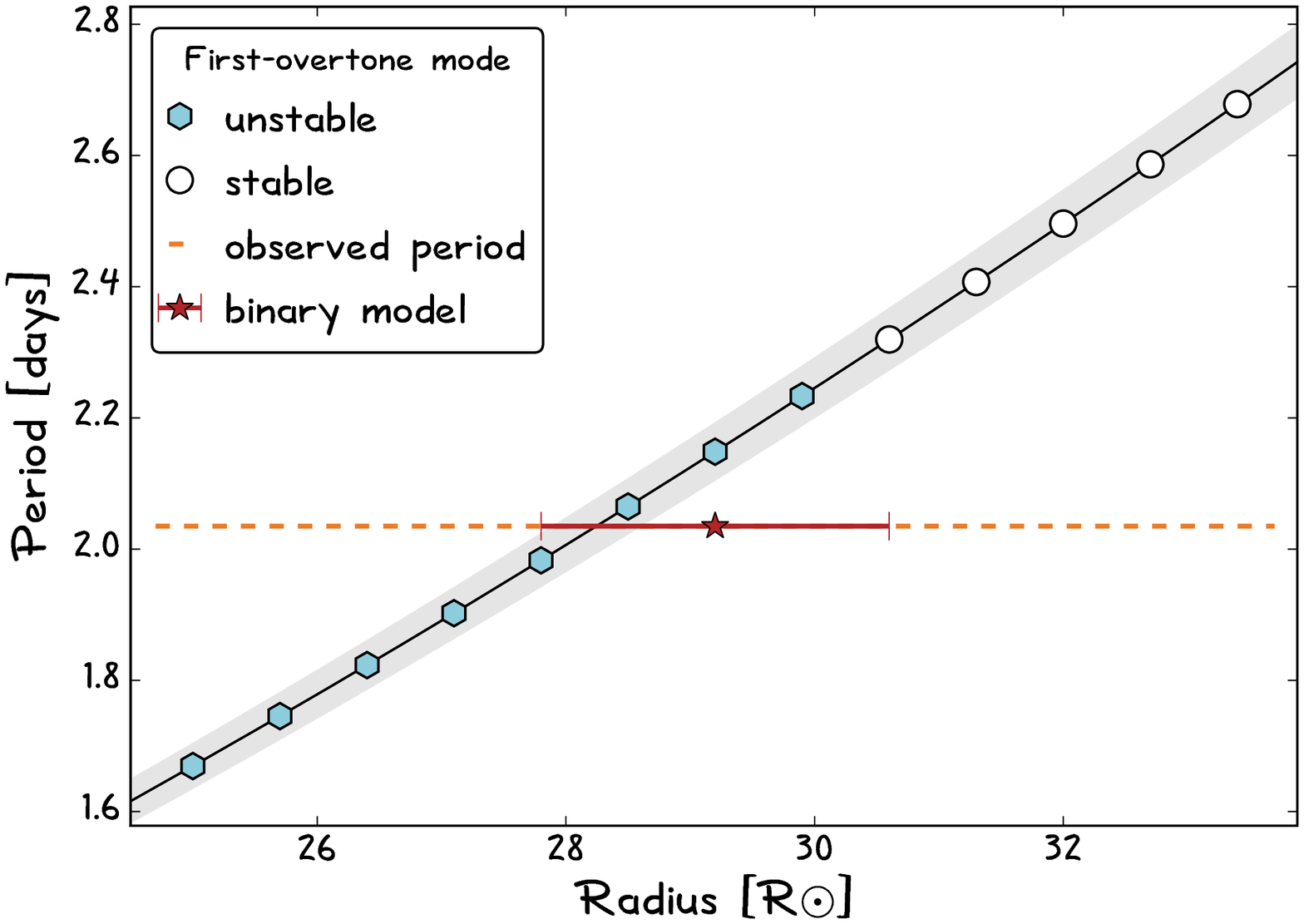}
    \caption{Calculated period vs.\ radius for LMC-CEP-2532. The radius determination from the eclipsing binary modeling can be improved using the pulsation theory.
    }
    \label{fig:fig2}
  \end{minipage}
\end{figure}

We collected the physical parameters of six Cepheids (LMC-CEP-0227, 1718, 1812, 2532, 4506) from the literature to compare the results with the pulsation theory models. For the computation of pulsation periods, we use the linear, non-adiabatic pulsation code of \cite{smolec:2008}.
We use the same model parameters as \cite{smolec:2012} and \cite{pilecki:2017}. In all model computations, we use the OPAL opacities, supplemented at low temperatures with the opacity data from \cite{ferguson:2005} and computed for the scaled solar chemical composition as given by \cite{asplund:2009}.
As input to compute the pulsation period for each Cepheid, we used their masses, radii, and effective temperatures together with the chemical composition ($X$, $Z$)
for five different metallicities, [Fe/H] = $0.0$, $-0.5$, $-1.0$, $-1.5$ and $-2.0$.
Depending on the pulsation mode of a given Cepheid, we obtain the computed fundamental or first-overtone mode period and their growth rates (see Fig.~\ref{fig:fig1}).

In general, the theoretical models are in very good agreement with the results from the observations. In the case of LMC-CEP-1718 and LMC-CEP-2532, only one eclipse per cycle is seen, and the radii from the binary system modeling are not very accurate. 
The pulsation models cannot reproduce the central values of the radii, but agree with the determined values within 1\,$\sigma$. As Fig.~\ref{fig:fig2} clearly shows, the pulsation models can improve the accuracy and precision of radii determination.

For LMC-CEP-1812, the small uncertainties of radius and mass from the binary model do not leave much space for the parameter adjustment in the pulsational model. A radius of 2\,$\sigma$ lower than the measured one is necessary for the calculated and observed periods to match.
As this Cepheid is a probable product of a merger \citep{neilson:2015}, the system may have a more complicated internal structure. This object needs further study.

\acknowledgements{We gratefully acknowledge financial support for this work from the Polish National Science Center, grant SONATA 2014/15/D/ST9/02248.}

\bibliographystyle{ptapap}
\bibliography{taormina_rrl2017}

\end{document}